PAPER

# Multifunctional composite magnet realizing record-high transverse thermoelectric generation

Fuyuki Ando,*[a] Takamasa Hirai,[a] Abdulkareem Alasli,[b] Hossein Sepehri-Amin,[a] Yutaka Iwasaki,[a] Hosei Nagano[b] and Ken-ichi Uchida*[ac]

Permanent magnets are used in various products and essential for human society. If omnipresent permanent magnets could directly convert heat into electricity, they would lead to innovative energy harvesting and thermal management technologies. However, achieving such "multifunctionality" has been difficult due to poor thermoelectric performance of conventional magnets. In this work, we develop a multifunctional composite magnet (MCM) that enables giant transverse thermoelectric conversion while possessing permanent magnet features. MCM comprising alternately and obliquely stacked $SmCo_5/Bi_{0.2}Sb_{1.8}Te_3$ multilayers exhibits an excellent transverse thermoelectric performance at room temperature by optimizing its anisotropic structure. Owing to the extremely low interfacial electrical and thermal resistivities, the experimentally determined figure of merit $z_{xy}T$ reaches 0.20 close to the analytically calculated ideal value. The MCM-based thermopile module generates 204 mW in maximum at a temperature difference of 152 K, whose power density normalized by heat transfer area and temperature gradient is not only record-high among transverse thermoelectric modules but also comparable to that of commercial thermoelectric modules utilizing the Seebeck effect. The multifunctionality of our MCM provides unprecedented opportunities for energy harvesting and thermal management everywhere permanent magnets are currently used.

## 1. Introduction

Transverse thermoelectric effects realize the interconversion between the charge and heat currents in the orthogonal direction. The orthogonal geometry simplifies the thermoelectric device architecture because it can eliminate substrates, electrodes, and their junctions in the thermal circuit. This junctionless structure makes it possible to apply a larger temperature gradient $\nabla T$ to thermoelectric materials owing to the absence of substrates, improve the thermoelectric conversion efficiency owing to the lack of interfacial thermal resistance between the thermoelectric materials and electrodes, and suppress thermal deterioration at hot-side junctions, all of which are significant issues for conventional longitudinal thermoelectric modules based on the Seebeck effect.[1–3] To utilize these geometrical advantages, many recent studies have focused on the development of physics, materials science, and device architectures for transverse thermoelectric conversion.

The transverse thermoelectric effects are classified into various mechanisms.[4] Among them, the four mechanisms are related to magnetism or spin: the ordinary and anomalous Nernst effects,[5–21] spin Seebeck effect,[22] and Seebeck-effect-driven anomalous Hall effect.[23] The other mechanisms are unrelated to magnetism or spin: the off-diagonal Seebeck effects (ODSEs)[3,24–40] in natural anisotropic crystals and artificial anisotropic composites. Recently, with the development of topological materials science and spin caloritronics, the anomalous Nernst effect (ANE) in magnetic materials has been intensively studied.[7–10,12–14,17,19] Thermoelectric conversion through ANE requires spatially uniform magnetization, which is typically achieved by applying external magnetic fields. From an application point of view, ANE for permanent magnets with remanent magnetization **M** has been studied to achieve magnetic-field-free operation of transverse thermoelectric conversion.[11,15,20] Although these developments realize multifunctionality enabling transverse thermoelectric conversion, the small transverse thermoelectric figure of merit $z_{xy}T$ for ANE ($<10^{-3}$) hinders the future applications of multifunctional magnets. By contrast, the studies on ODSE have independently progressed and predicted a considerably higher $z_{xy}T$ ($>0.1$ at room temperature) compared with that for ANE.[28,33,34,37–39] ODSE is enhanced when the two materials having opposite Seebeck coefficients, i.e., *p*- and *n*-type materials, and large differences in electrical and thermal conductivities are alternately and obliquely stacked (artificially tilted multilayers, ATML).[27] However, despite the wide material selectivity, no attempts have been made to integrate the magnetic functionality into ATMLs, except for a recent study.[40] Although ATMLs consisting of $Nd_2Fe_{14}B$-type permanent magnets and $Bi_{88}Sb_{12}$ have been developed, their transverse thermoelectric performance is poor ($z_{xy}T < 2.0 \times 10^{-3}$) because the sign of the Seebeck coefficient of $Nd_2Fe_{14}B$ is the same as that of $Bi_{88}Sb_{12}$.

[a] National Institute for Materials Science, Tsukuba 305-0047, Japan
[b] Department of Mechanical Systems Engineering, Nagoya University, Nagoya 464-8603, Japan
[c] Department of Advanced Materials Science, Graduate School of Frontier Sciences, The University of Tokyo, Kashiwa 277-8561, Japan

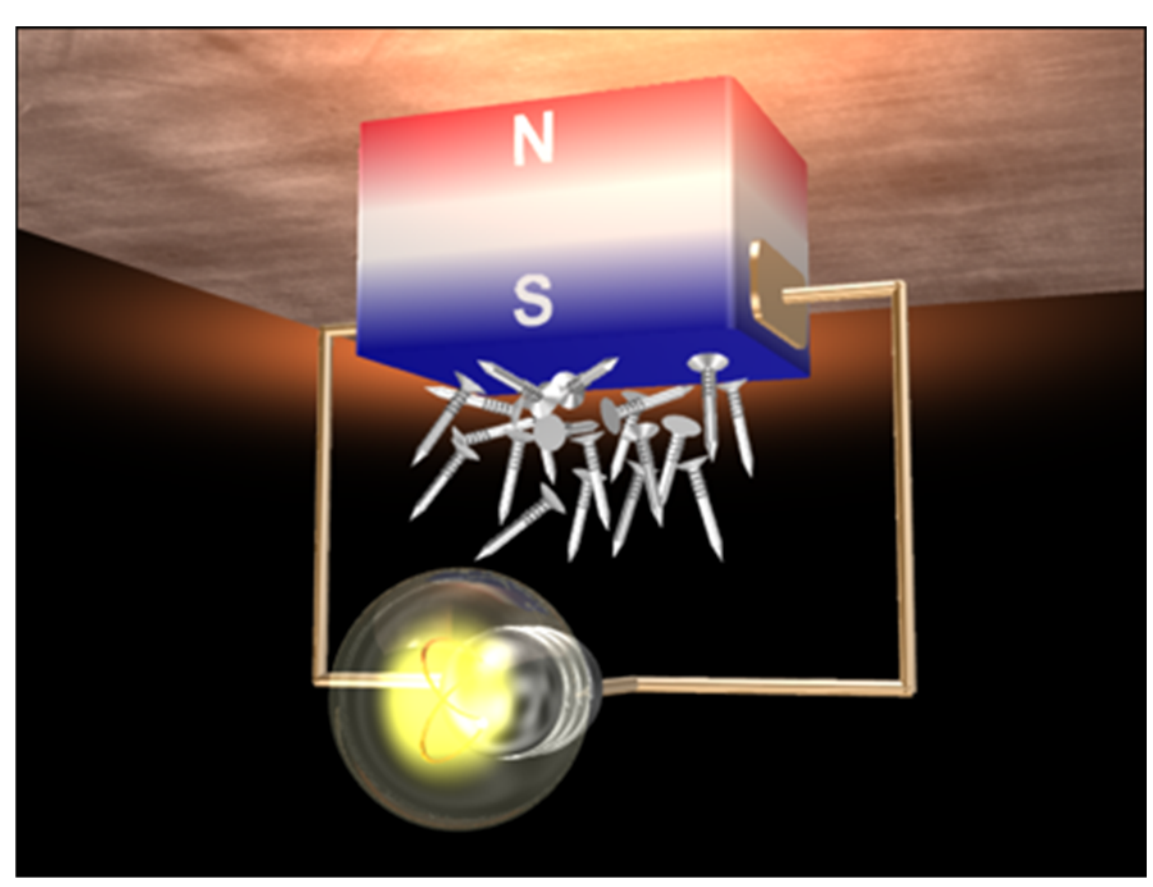

**Fig.1** Schematic of the concept of MCM exhibiting strong magnetic force and superior transverse thermoelectric performance simultaneously.

Here, we have developed MCM that exhibits giant transverse thermoelectric conversion in addition to large remanent magnetization and coercivity (Fig. 1). Our MCM, comprising alternately and obliquely stacked $SmCo_5$-type permanent magnets ($SmCo_5$) and thermoelectric $Bi_{0.2}Sb_{1.8}Te_3$ (BST) slabs, experimentally exhibits an excellent $z_{xy}T$ of 0.20 at room temperature owing to the optimized anisotropic composite structure and extremely low interfacial electrical and thermal resistivities between the $SmCo_5$ and BST layers. Utilizing these high-performance MCM elements, we constructed a lateral thermopile module to obtain a higher thermoelectric voltage while maintaining the magnetic functionality. The MCM-based module exhibits an output power $P$ of 204 mW at a temperature difference $\Delta T$ of 152 K, which corresponds to the record-high normalized power density per heat transfer area and $\nabla T^2$ of 0.17 $mW/cm^2 \cdot (mm/K)^2$ among the transverse thermoelectric modules.

## 2. Results and discussion

### 2.1 Magnetic and analytical transverse thermoelectric properties

ODSE in ATMLs generates of an electric field $\mathbf{E}$ in the direction perpendicular to $\nabla T$ owing to the anisotropic transport properties in the tilted axes with respect to the $\nabla T$ direction (Fig. 2a). Due to the difference in the thermal conductivity of the constituent materials, the heat current $\mathbf{J}_q$ nonuniformly flows in ATMLs by the application of $\nabla T$. As a result of thermoelectric conversion in the $p$- and $n$-type materials, the transverse component of the charge current $\mathbf{J}_c$ is additively generated (see Supplementary Note S1 and Fig. S1, ESI†, where the nonuniform charge-to-heat current conversion as the origin of ODSE was confirmed by the lock-in thermography method). Here, the primary characteristic of ODSE is that, even if the Seebeck coefficient $S$ of the constituent material is not high, $z_{xy}T$ can be enhanced by combining $p$- and $n$-type materials whose electrical and thermal conductivities ($\sigma$ and $\kappa$) are largely different. Indeed, ATMLs composed of metals and thermoelectric semiconductors, such as $n$-type Ni/$p$-type $Bi_{0.5}Sb_{1.5}Te_3$,[28,30,31,34] $n$-type $YbAl_3$/$p$-type $Bi_{0.5}Sb_{1.5}Te_3$,[33] $p$-type Fe/$n$-type $Bi_2Te_{2.7}Se_{0.3}$,[37] and $n$-type Co/$p$-type $Bi_{0.5}Sb_{1.5}Te_3$,[38] have been predicted to exhibit $z_{xy}T > 0.1$ at room temperature as well as $n$-type $Bi_2Te_{2.7}Se_{0.3}$/$p$-type $Bi_{2-x}Sb_xTe_3$.[39] $SmCo_5$, which is a widely known permanent magnet for its strong magnetic force and excellent thermal stability,[41] also has metallic transport properties and negative $S$.[11,15] Thus, we selected $p$-type BST as a counterpart material for $SmCo_5$, as $\sigma$ and $\kappa$ are one order smaller than those of $SmCo_5$ and the sign of $S$ is opposite (Fig. S2, ESI†).

We predict the superior transverse thermoelectric performance in $SmCo_5$/BST-based ATML using analytical matrix calculations. Based on Goldsmid's method,[27] the following transverse thermoelectric properties can be calculated with neglecting interfacial contributions: the electrical conductivity in the $x$-axis $\sigma_{xx}$, thermal conductivity in the $y$-axis $\kappa_{yy}$, off-diagonal Seebeck coefficient $S_{xy}$, and resultant $z_{xy}T$ (= $S_{xy}^2\sigma_{xx}T/\kappa_{yy}$). The transport parameters of $SmCo_5$/BST multilayers in the direction parallel ($\sigma_{\parallel}$, $\kappa_{\parallel}$, and $S_{\parallel}$) and perpendicular ($\sigma_{\perp}$, $\kappa_{\perp}$, and $S_{\perp}$) to the stacking plane are analytically calculated using the electrical conductivities $\sigma_{SmCo}$ and $\sigma_{BST}$, thermal conductivities $\kappa_{SmCo}$ and $\kappa_{BST}$, and Seebeck coefficients $S_{SmCo}$ and $S_{BST}$ for $SmCo_5$ and BST, respectively:[27]

$$\sigma_{\parallel} = t\sigma_{SmCo} + (1-t)\sigma_{BST} \qquad \sigma_{\perp} = \frac{\sigma_{SmCo}\sigma_{BST}}{(1-t)\sigma_{SmCo} + t\sigma_{BST}} \tag{1}$$

$$\kappa_{\parallel} = t\kappa_{SmCo} + (1-t)\kappa_{BST} \qquad \kappa_{\perp} = \frac{\kappa_{SmCo}\kappa_{BST}}{(1-t)\kappa_{SmCo} + t\kappa_{BST}} \tag{2}$$

$$S_{\parallel} = \frac{t\sigma_{SmCo}S_{SmCo} + (1-t)\sigma_{BST}S_{BST}}{t\sigma_{SmCo} + (1-t)\sigma_{BST}} \qquad S_{\perp} = \frac{t\kappa_{BST}S_{SmCo} + (1-t)\kappa_{SmCo}S_{BST}}{t\kappa_{BST} + (1-t)\kappa_{SmCo}} \tag{3}$$

where $t = d_{SmCo}/(d_{SmCo}+d_{BST})$ denotes the thickness ratio of the $SmCo_5$ layer and $d_{SmCo}$ ($d_{BST}$) the thickness of the $SmCo_5$ (BST) layer. When the homogeneous $SmCo_5$/BST multilayers are rotated in the $xy$-plane with the tilt angle $\theta$ (Fig. 2a), the transverse thermoelectric properties by ODSE can be expressed as

$$\sigma_{xx} = \frac{\sigma_{||}\sigma_{\perp}}{\sigma_{||}\sin^2\theta + \sigma_{\perp}\cos^2\theta} \tag{4}$$

$$\kappa_{yy} = \kappa_{||}\sin^2\theta + \kappa_{\perp}\cos^2\theta \tag{5}$$

$$S_{xy} = \left(S_{\perp} - S_{||}\right)\sin\theta\cos\theta \tag{6}$$

Then, the transverse thermoelectric figure of merit $z_{xy}T$ is given by

$$z_{xy}T = \frac{{S_{xy}}^2\sigma_{xx}}{\kappa_{yy}}T \tag{7}$$

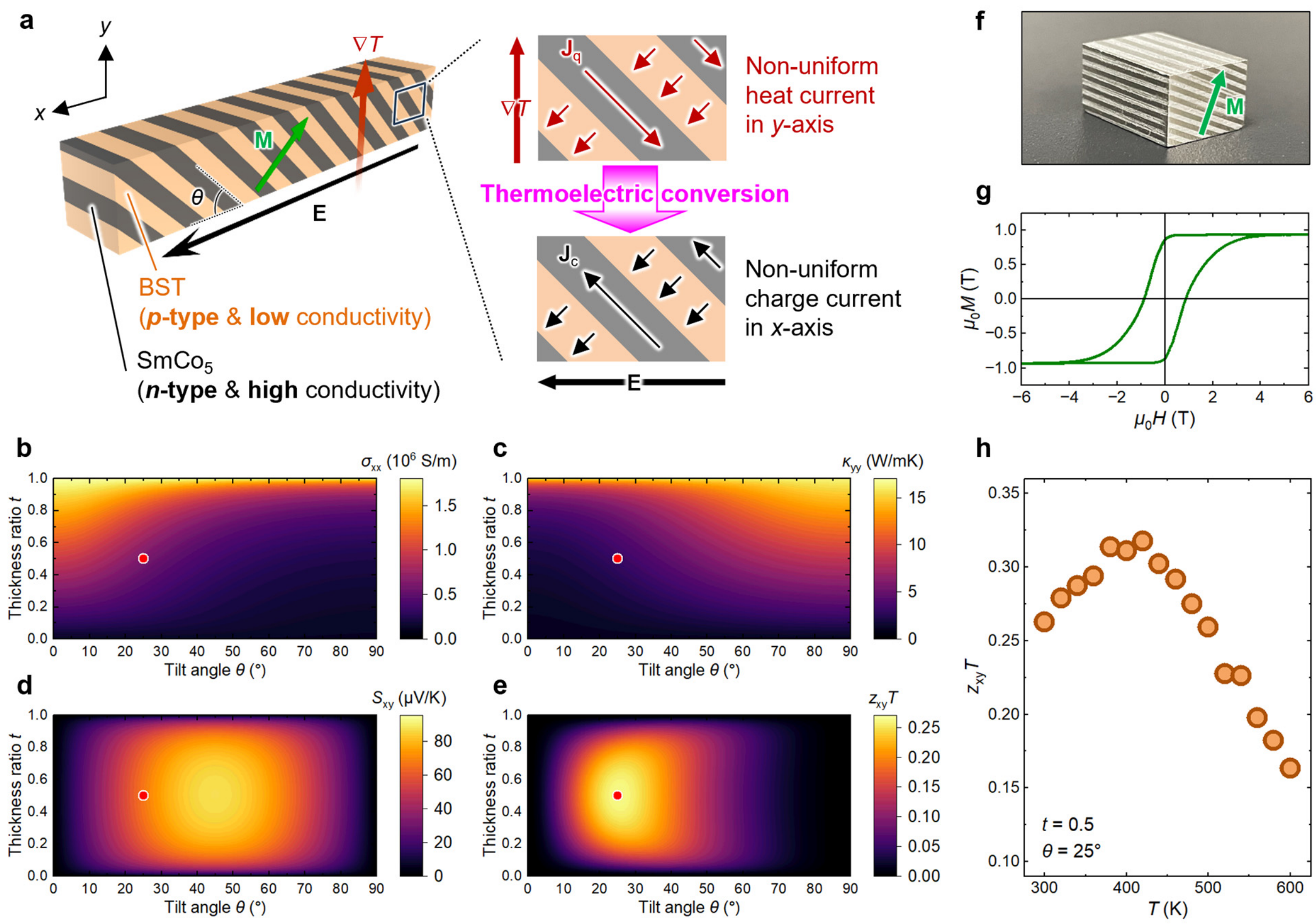

**Fig.2** Magnetic and analytical transverse thermoelectric properties of MCM. (a) Schematic of MCM composed of $SmCo_5$ and BST with a tilt angle $\theta$. An ATML structure induces a nonuniform heat current $\mathbf{J}_q$ by the application of a temperature gradient $\nabla T$ in the $y$-axis, which generates a charge current $\mathbf{J}_c$ in the transverse direction through thermoelectric conversion by the Seebeck effect. Remanent magnetization in the $SmCo_5$ layers $\mathbf{M}$ is oriented along the stacking direction. (b–e) Contour plots of the analytical electrical conductivity in the $x$-axis $\sigma_{xx}$ (b), thermal conductivity in the $y$-axis $\kappa_{yy}$ (c), off-diagonal Seebeck coefficient $S_{xy}$ (d), and transverse thermoelectric figure of merit $z_{xy}T$ (e) as functions of the thickness ratio of the $SmCo_5$ layer $t$ and $\theta$ in $SmCo_5$/BST-based MCM at 300 K. Red points indicate the optimum $t$ and $\theta$ to maximize $z_{xy}T$, i.e., $t$ = 0.5 and $\theta$ = 25°. (f) Photograph of $SmCo_5$/BST-based MCM. (g) Magnetization $M$ of the $SmCo_5$ slab as a function of an external magnetic field $H$ in its magnetic easy axis direction. $\mu_0$ is the vacuum permeability. (h) Temperature $T$ dependence of the calculated $z_{xy}T$ at $t$ = 0.5 and $\theta$ = 25°.

Fig. 2b–e shows the thickness ratio $t$ and tilt angle $\theta$ dependences of the transverse thermoelectric properties for $SmCo_5$/BST-based ATML, obtained by substituting the measured properties of $SmCo_5$ and BST into eqn (1–7). We found that the calculated $z_{xy}T$ in $SmCo_5$/BST-based ATML reaches 0.26 at the optimum $t$ of 0.5 and $\theta$ of 25° at $T$ = 300 K (see red points in Fig. 2b–e). Furthermore, we calculated the temperature $T$ dependence of $\sigma_{xx}$, $\kappa_{yy}$, $S_{xy}$, and $z_{xy}T$ in the range of 300–600 K at $t$ = 0.5 and $\theta$ = 25° (Fig. 2h and Fig. S3, ESI†). The $z_{xy}T$ value reaches 0.32 at 420 K in maximum, which is more than two orders of magnitude higher than that of $Nd_2Fe_{14}B$/BiSb-based ATML.[40]

To experimentally demonstrate the expected performance as MCM, we synthesized $SmCo_5$/BST-based ATML using the calculated optimum $\theta$ and $t$ values. $SmCo_5$ circular disks and BST powders were alternately stacked and bonded using spark plasma sintering, followed by cutting the sintered multilayer into tilted rectangular blocks (see Fig. 2f and Experimental section for details). The accuracies of $t$ and $\theta$ in the ATML block were estimated to be 0.50 ± 0.05 and 25 ± 1°, respectively. We observed the elemental distribution maps of Sm, Co, Bi, Sb, and Te measured using scanning electron microscopy with energy dispersive X-ray spectroscopy (SEM-EDX). The low magnification EDX mapping of Sb, Te, and Co and the atomic ratio profile indicate no complex elemental migrations between the $SmCo_5$ and BST layers (Fig. S4, ESI†). Then, from the high magnification image around the $SmCo_5$/BST interface shown in Fig. 3a, we recognize the growth of interfacial diffusion layers with a thickness of approximately 10 μm without interfacial voids. The atomic ratio profile in the stacking direction of the $SmCo_5$ and BST multilayers (Fig. 3b) reveals that the interfacial diffusion layers of $CoTe_a$-based compounds ($1.5 < a < 2.0$) are formed, which act as adhesive bonds. The magnetic easy axis of the $SmCo_5$ disks was out-of-plane. Fig. 2g shows the magnetization $M$ of the $SmCo_5$ portion cut from ATML as a function of the magnetic field $H$ perpendicular to the stacking plane. The $SmCo_5$ layers exhibit large remanent magnetization of 0.86 T and coercivity of 0.87 T without deterioration even after sinter-bonding with the BST layers, which confirms that $SmCo_5$/BST-based ATML operates as MCM.

### 2.2 Contribution of interfacial electrical and thermal resistivities

ODSE in ATMLs generates of an electric field $\mathbf{E}$ in the direction perpendicular to To verify the transverse thermoelectric performance of $SmCo_5$/BST-based MCM, it is important to characterize the interfacial electrical and thermal transport properties at the $SmCo_5$/BST junctions because they cause a degradation from the calculated properties in Fig. 2b–e as well as the thermoelectric material/electrode junctions in longitudinal thermoelectric devices.[2,42–44] Despite their importance, the quantitative investigation of the interfacial transport properties has not been attempted in previous studies.[30,33,34,37,38,40] In this study, we experimentally characterized the interfacial electrical and thermal resistivities and their contributions to the volumetric resistances.

First, we characterized the interfacial electrical resistance between the $SmCo_5$ and BST layers by measuring the spatial distribution of the electrical resistance. To distinguish the resistances originating from the bulk of $SmCo_5$, bulk of BST, and their interfaces, we prepared a rectangular sample comprising the $SmCo_5$/BST multilayer in which the cut angle was perpendicular to the stacking plane (i.e., $\theta = 90°$). As shown in Fig. 3c, the four-terminal ac resistance was measured while the probe was scanned along the stacking direction. The measurement results shown in Fig. 3d revealed a step-like behavior reflecting the different electrical resistivities of $SmCo_5$ and BST layers, wherein the electrical resistivity of $SmCo_5$ is much smaller than that of BST (Fig. S2, ESI†). Significantly, the resistance profile was observed to be continuous at the interfaces, the positions of which are indicated by gray dotted lines in Fig. 3d. The interfacial electrical resistivity was estimated to be 0.4 ± 1.2 $\mu\Omega cm^2$ by averaging the resistance gap values at several interfaces by linearly fitting the resistance profiles in the $SmCo_5$ and BST layers and extrapolating the fitting functions to the interface positions. This unusually low interfacial resistivity, comparable to lowest-level contact resistivities for thermoelectric devices,[42] can be attributed to the metallic property of $CoTe_a$-based diffusion layers,[45–47] whose conductivity is comparable to that of $SmCo_5$. Meanwhile, the volume resistance-area product of the 0.5-mm-thick $SmCo_5$ and BST layers was estimated to be 64 ± 2 $\mu\Omega/cm^2$ from the slopes of the fitting functions. Thus, the ratio of the interfacial electrical resistance to the volumetric resistance was 1.2%, which is negligibly small within the margin of the experimental error.

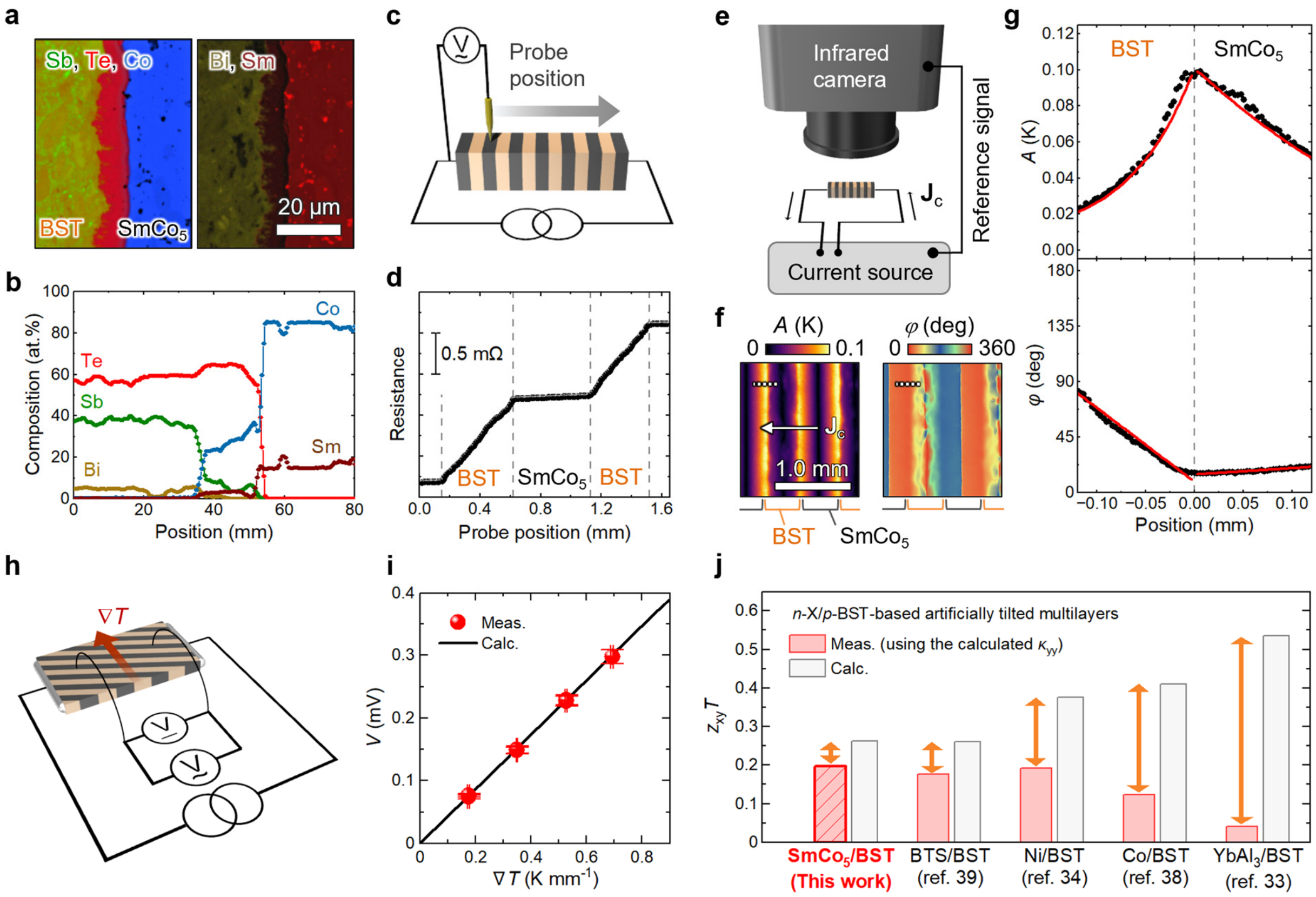

**Fig.3** Contribution of interfacial transport properties to the transverse thermoelectric performance of MCM. (a) SEM-EDX image of the cross section of the $SmCo_5$/BST multilayer. (b) Line profile of the atomic ratio of Sm, Co, Bi, Sb, and Te across the stacking direction. (c) Schematic of the four-terminal resistance measurement setup to investigate the spatial distribution of the electrical resistance by sweeping the probe position. (d) Probe position dependence of the electrical resistance in the $SmCo_5$/BST multilayer in the stacking direction. Gray dotted lines indicate the approximate positions of the $SmCo_5$/BST interfaces. (e) Schematic of the LIT measurement while applying the square-wave-modulated charge current $\mathbf{J}_c$ with the amplitude $J_c$ and frequency $f$. (f) Lock-in amplitude $A$ and phase $\varphi$ images of the cross section of the $SmCo_5$/BST multilayer at $J_c$ = 1 A and $f$ = 25 Hz. (g) Line profiles of $A$ and $\varphi$ across the $SmCo_5$/BST interface for the white dotted lines with a length of 101 pixels in (f). Red curves represent the fitting functions using the one-dimensional heat equation. (h) Schematic of the setup for the direct measurement of $S_{xy}$ and $\sigma_{xx}$. (i) The temperature gradient $\nabla T$ dependence of the directly measured and analytically calculated thermoelectric voltage $V$ values at room temperature. (j) Comparison of the measured and calculated $z_{xy}T$ at room temperature for various $n$-type X/$p$-type BST-based ATMLs (X = $SmCo_5$, $Bi_2Te_{2.7}Se_{0.3}$, Ni, Co, and $YbAl_3$). The calculated $\kappa_{yy}$ values are used for both the measured and calculated $z_{xy}T$.

The interfacial thermal resistance was also characterized using the lock-in thermography (LIT) measurements[48–50] for the same $SmCo_5$/BST multilayer sample with $\theta = 90°$. Fig. 3e shows a schematic of the LIT measurement setup. A square-wave-modulated $\mathbf{J}_c$ with an amplitude $J_c$ of 1 A and a frequency $f$ of 25 Hz was applied to the sample in a direction perpendicular to the stacking plane. The thermal images were continuously captured while applying $\mathbf{J}_c$ to observe the temperature modulation due to the Peltier-effect-induced heat current. When the heat current is discontinuous due to the difference in the Peltier coefficient at junctions, finite heat absorption and release appear in the vicinity of the interfaces.[49,50] By extracting the first harmonic response of the charge-current-induced temperature modulation through Fourier analysis and calculating the lock-in amplitude $A$ and phase $\varphi$ for each pixel of the thermal images, we visualized the pure contribution of the Peltier effect without contamination by Joule heating. Here, the $A$ signals refer to the magnitude of the temperature change in linear response to $\mathbf{J}_c$ and $\varphi$ signals to its sign and time delay due to the heat diffusion. A previous study[50] reported that the spatial profiles of the $A$ and $\varphi$ signals can be used to investigate the interfacial thermal resistance because a finite interfacial thermal resistance causes discontinuities in $A$ and $\varphi$. As shown in Fig. 3f, we observed clear $A$ signals near the $SmCo_5$/BST interfaces and $\varphi$ signals alternately changing from approximately 0° to 180° for each adjacent interface, which is consistent with the features of the Peltier-effect-induced temperature modulation. The line profiles of $A$ and $\varphi$ across the $SmCo_5$/BST interface are shown in Fig. 3g, where no obvious jumps appear at the interface position. By fitting the position dependence of $A$ and $\varphi$ using the one-dimensional heat equation,[50] the interfacial thermal resistance was estimated to be less than $1\times10^{-6}$ $m^2K/W$. Meanwhile, the thermal conductivities of $SmCo_5$ and BST layers were 16.3 and 1.0 W/mK at 300 K, respectively (Fig. S2, ESI†). Thus, the contribution of the interfacial thermal resistance to the volumetric thermal resistance was also negligible (<0.4%).

The above experiments ensure that the performance degradation due to the presence of multiple interfaces is quite small in our $SmCo_5$/BST-based MCM. To experimentally determine $z_{xy}T$ with negligible contributions from the interfacial electrical and thermal resistances, we directly measured $S_{xy}$ and $\sigma_{xx}$ in $SmCo_5$/BST-based MCM with $\theta = 25°$ at room temperature. Fig. 3h shows the measurement setup for the transverse thermoelectric voltage $V$ induced by a temperature gradient $\nabla T$ and the four-terminal ac resistance (see Experimental section for details). Fig. 3i shows that the $V$ value linearly increases with $\nabla T$ and quantitatively agrees with the calculated line obtained from the analytically calculated $S_{xy}$ (Fig. 2d), indicating that our MCM exhibits the ideally large transverse thermoelectric conversion. The $S_{xy}$ value of $SmCo_5$/BST-based MCM was experimentally estimated to be 66.4 ± 1.1 μV/K at room temperature. The ac resistance was measured to be 1.54 ± 0.02 mΩ while the calculated one is 1.17 mΩ, where the increased resistance can be attributed to cracking in the $SmCo_5$ disks during the SPS process and to processing errors, such as $\theta$ and $t$.

Fig. 3j shows the comparison of $z_{xy}T$ between the values estimated from the analytical matrix calculation and direct measurement of $S_{xy}$ and $\sigma_{xx}$ for various $n$-type X/$p$-type BST-based ATMLs, where X = $SmCo_5$, $Bi_2Te_{2.7}Se_{0.3}$ (BTS), Ni, Co, and $YbAl_3$.[33,34,38,39] Note that the analytical $\kappa_{yy}$ values are used for both calculated and measured $z_{xy}T$ because the direct measurement of $\kappa_{yy}$ has never been done due to the difficulty to exclude the contaminated boundary effect. In the previous reports, the higher measured electrical resistances predominantly caused degradations in $z_{xy}T$ from the calculated ones, the magnitude of which varies depending on the materials combination. Thus, even if the analytical calculation suggests the higher $z_{xy}T$ values, the actual $z_{xy}T$ value can be lower due to the interfacial contributions ($z_{xy}T$ < 0.2 when the analytically calculated $\kappa_{yy}$ is used). On the contrary, although the calculated $z_{xy}T$ value of our $SmCo_5$/BST-based MCM is not the best, the extremely low interfacial resistances between the $SmCo_5$ and BST layers successfully suppress the performance degradation and realize the high $z_{xy}T$ value of 0.20 at room temperature. Thus, the introduction of $SmCo_5$ as the counterpart of BST not only provides the magnetic functionality but also contributes to the record-high transverse thermoelectric performance.

### 2.3 Giant transverse thermoelectric generation in MCM-based module

In this section, we demonstrate giant transverse thermoelectric generation using $SmCo_5$/BST-based MCM. To achieve this, we developed a lateral thermopile module to enhance the thermoelectric voltage for a useful power supply to drive widely-used electronic devices. Fig. 4a shows a schematic of the module structure, where the thin-sliced MCM elements are stacked with an opposite $\theta$ between the neighboring elements and intermediated by thin insulator layers. By electrically connecting the ends of the neighboring MCMs, a long series circuit is formed so that $\mathbf{J}_c$ flows in a zigzag manner, as proposed by Norwood[1] and Kanno.[30] Importantly, the net magnetization of the thermopile module $\mathbf{M}_{net}$ is oriented along the $\nabla T$ direction by the vector sum of **M** in each MCM element. Thus, the MCM-based thermopile module operates as a permanent magnet. Fig. 4b shows the photograph of the constructed module, which is composed of 14 elements of magnetized $SmCo_5$/BST-based MCM (see Experimental section for details). The fill factor of the MCM elements per heat transfer area exceeds 90% owing to the transverse geometry and very thin (approximately 0.05-mm-thick) insulator layers, ensuring a high thermoelectric output density and robustness against mechanical stress.[2,3] As shown in Fig. 4b, several ferromagnetic metal paper clips can be hung to the magnetized MCM-based module owing to its large $\mathbf{M}_{net}$. The internal resistance of the MCM-based module $R_{module}$ was measured to be 34.8 mΩ by the four-terminal measurement, which deviates from the calculated $R_{module}$ only by +11%. Thus, the contribution of the contact resistance between the MCM elements to the total resistance is also small. Fig. 4c shows a schematic of the measurement setup for transverse thermoelectric generation. The MCM-based module was sandwiched by the heater and heat bath to apply $\nabla T$. The four-terminal voltage and power measurements were performed in the MCM-based module in the demagnetized state. The reason why the magnetic state (**M** and its stray field) of $SmCo_5$ does not contribute to the transverse thermoelectric performance is explained in Supplementary Note S2 (ESI†).

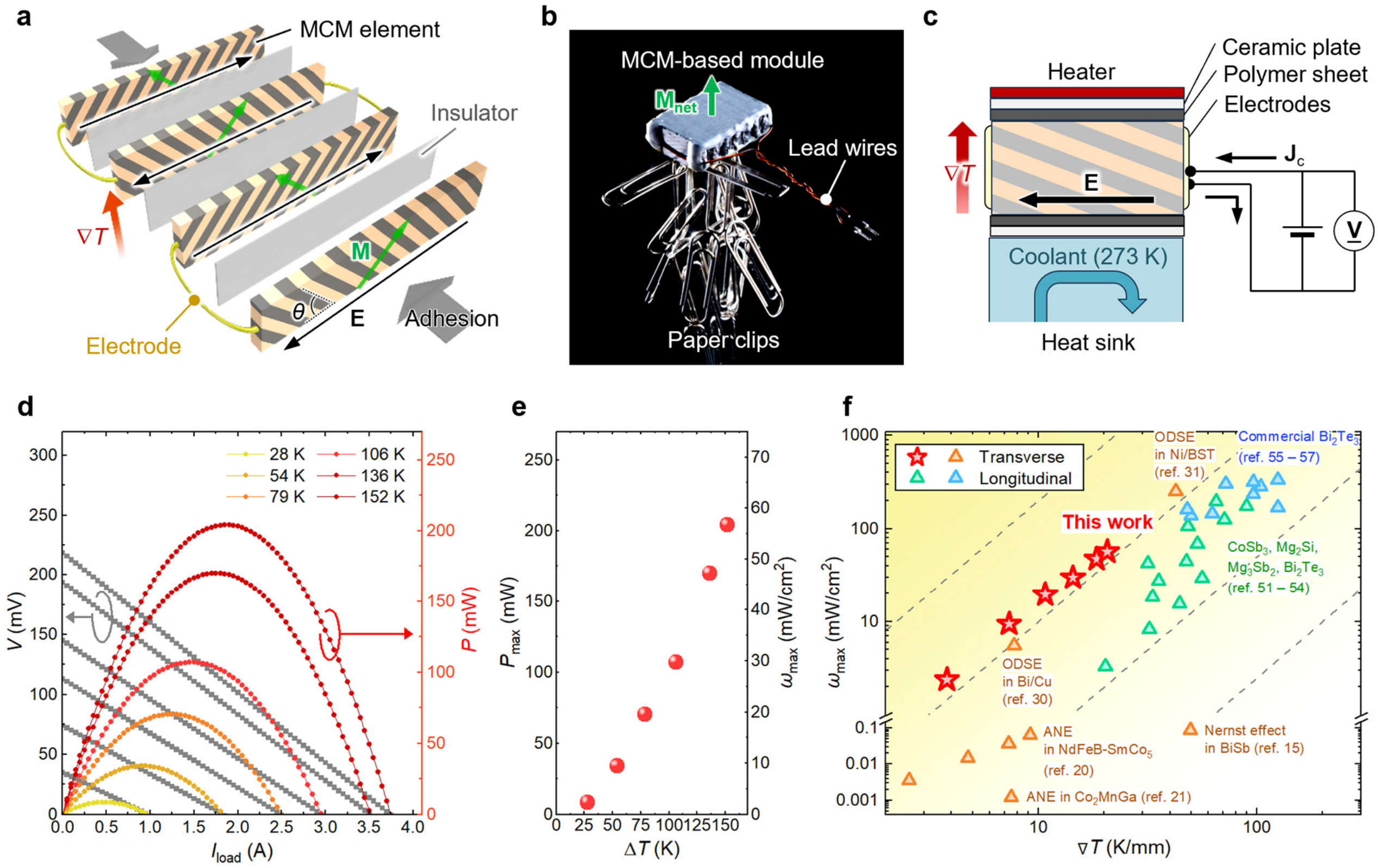

**Fig. 4** Transverse thermoelectric generation by MCM-based thermopile module. (a) Schematic of the architecture of the MCM-based thermopile module. The MCM elements are alternately stacked with opposite $\theta$ intermediated by thin insulator layers. By attaching electrodes to connect the ends of neighboring MCMs in a zigzag manner, a long series circuit is formed to sum up the transverse thermoelectric voltage. Net magnetization of the thermopile module $\mathbf{M}_{net}$ is oriented along the $\nabla T$ direction by the vector sum of $\mathbf{M}$ in each MCM element. (b) Photograph of the magnetized MCM-based module hanging many ferromagnetic metal paper clips owing to the large $\mathbf{M}_{net}$. Lead wires are connected at the ends of the thermopile circuit. (c) Schematic of the measurement setup for the transverse thermoelectric generation. (d) Load current $I_{load}$ dependence of the thermoelectric voltage $V$ and output power $P$ at various values of the temperature difference $\Delta T$. (e) $\Delta T$ dependence of the maximum output power $P_{max}$ and maximum power density per heat transfer area $\omega_{max}$. (f) $\nabla T$ dependence of $\omega_{max}$ for the MCM-based module used in this study (red stars), transverse (orange triangles) and longitudinal (green triangles) thermoelectric modules reported in the research papers, and commercial longitudinal thermoelectric modules (blue triangles). Gray dotted lines represent functions of $\omega_{max} = a\nabla T^2$ with $a$ = 0.00025, 0.005, 0.1, and 2.

Fig. 4d,e shows the results of the transverse thermoelectric generation in the MCM-based module. In the load current $I_{load}$ dependence of the thermoelectric voltage $V$ (Fig. 4d), the open-circuit voltage $V_{oc}$ is defined as the $V$ value at $I_{load}$ = 0 A and $V$ linearly decreases with increasing $I_{load}$ according to the internal resistance. $P$ (= $I_{load}V$) shows a parabolic behavior against $I_{load}$ and has a maximum value ($P_{max}$) when $V$ is the half of $V_{oc}$. The $V_{oc}$ and $P_{max}$ values monotonically increased with increasing $\Delta T$

and reached 219 mV and 204 mW, respectively, at $\Delta T$ = 152 K, which are giant values for the transverse thermoelectric generation owing to the excellent $z_{xy}T$ with low interfacial electrical and thermal resistivities. Fig. 4e shows the $\Delta T$ dependence of $P_{max}$ and the corresponding maximum power density $\omega_{max}$ per heat transfer area, which were almost parabolically increased with increasing $\Delta T$ as they are proportional to the square of $V_{oc}$, indicating almost no thermal deterioration of the thermoelectric properties at large $\Delta T$. Consequently, $\omega_{max}$ reached 56.7 mW/cm$^2$ at $\Delta T$ = 152 K owing to the transverse thermoelectric properties of MCM and the high fill factor of >90%. From the $P_{max}$ value and the analytical transverse thermoelectric properties, we also estimated the conversion efficiency for our MCM-based thermopile module at $\Delta T$ = 152 K to be 1.6–2.4% (Supplementary Note S3 and Fig. S5, ESI†).

We compare the thermoelectric power generation performance of our MCM-based module with that of various transverse and longitudinal thermoelectric modules including commercial products. To fairly compare the intrinsic performance, we show the $\nabla T$ dependence of $\omega_{max}$ in Fig. 4f because the $\omega_{max}$ value is proportional to the square of $\nabla T$ and independent of the mechanism, geometry, and size of the thermoelectric modules. Gray dotted guide lines in Fig. 4f represent functions of $\omega_{max} = a\nabla T^2$ with $a$ = 0.00025, 0.005, 0.1, and 2. Transverse thermoelectric modules utilizing the ordinary Nernst effect and ANE based on BiSb, $Co_2MnGa$, and $Nd_2Fe_{14}B/SmCo_5$ exhibit $\omega_{max}$ less than 0.1 mW/cm$^2$ regardless of $\nabla T$ typically due to the low thermopower,[16,20,21] which has been the barrier towards applications of transverse thermoelectrics. Meanwhile, the transverse thermoelectric modules composed of Bi/Cu- and Ni/BST-based ATMLs reported higher $\omega_{max}$ of 5.5 and 250 mW/cm$^2$ at $\nabla T$ = 7.8 and 42.5 K/mm, respectively.[30,31] However, the problem of the conventional ODSE-based module is the performance degradation due to the interfacial resistances. In this study, by synthesizing high-performance MCM with extremely low interfacial resistances and constructing a high-density thermopile structure, we successfully demonstrated $\omega_{max}$ of 56.7 mW/cm$^2$ at $\nabla T$ = 20.7 K/mm in maximum (the red stars in Fig. 4f). Here, let us compare the normalized power density $\omega_{max}/\nabla T^2$ in the similar temperature range to exclude the contribution of $\nabla T$. Then, our MCM-based module shows the record-high power generation performance of $\omega_{max}/\nabla T^2$ = 0.17 mW/cm$^2$·(mm/K)$^2$ among all the transverse thermoelectric modules including Bi/Cu- and Ni/BST-based ATMLs (0.09 and 0.14 mW/cm$^2$·(mm/K)$^2$, respectively). In Fig. 4f, we also show the $\omega_{max}$ values of the longitudinal thermoelectric modules utilizing the Seebeck effect, based on $CoSb_3$, $Mg_2Si$, $Mg_3Sb_2$, and $Bi_2Te_3$, including commercial products. The $\omega_{max}$ values are in the range of 15–497 mW/cm$^2$ at $\Delta T$ of 120–200 K,[51–57] and the corresponding $\omega_{max}/\nabla T^2$ values are calculated to be less than 0.07 mW/cm$^2$·(mm/K)$^2$. Surprisingly, the potential thermoelectric power density of our MCM-based module is larger than that of the commercial longitudinal thermoelectric modules. Thus, while having versatile transverse geometry and high mechanical durability,[1,7] our MCM-based module can generate the practical-level thermoelectric output power.

## 3. Discussion

Herein, we discuss the possible future developments of MCMs in terms of the power generation performance. While the experimentally determined $z_{xy}T$ value of 0.20 for our MCM and the estimated conversion efficiency of 1.6–2.4% at $\Delta T$ = 152 K for our MCM-based thermopile module are record-high among the transverse thermoelectric modules, the thermoelectric performance can be further improved through the following two approaches. One is the direct way to enhance ODSE by exploring permanent magnet materials with higher $S$ and $\sigma$ than those of commercial $SmCo_5$-type magnets and optimizing interfacial microstructures with low electrical and thermal resistivities, as detailed in this work. Another approach involves hybridizing other mechanisms related to magnetism or spin. We have previously demonstrated the hybridization of the off-diagonal Peltier effect with the magneto-Peltier, ordinary Ettingshausen effects, and ANE in ATMLs,[40,58] where the calculated $z_{xy}T$ varies from 0.14 to 0.20 depending on an external magnetic field. In this study, we did not introduce such hybrid thermoelectric conversion to prioritize superior multifunctionality. However, it is possible to improve $z_{xy}T$ by hybridizing ANE and the Seebeck-effect-driven anomalous Hall effect in permanent magnets and/or the magneto-Seebeck effect in thermoelectric materials.[6,23,59] Thus, exploring magnetic materials with large anomalous Nernst and Hall effects is essential for further development of MCMs. Meanwhile, from the technological point of view, the magnetic attractive force enables an efficient heat input/output by reducing the contact thermal resistances between MCM and a heat source/bath made of a magnetic material. Supplementary Note S4 and Fig. S6 (ESI†) show a proof-of-concept demonstration of energy harvesting using the MCM-based module from a heat source made of a steel use stainless, which suggests the usefulness of the magnetic functionality.

## 4. Conclusions

We developed a novel functional material named MCM and demonstrated giant transverse thermoelectric generation in the MCM-based thermopile module. Our experiments showed that MCM consisting of alternately and obliquely stacked $SmCo_5$/BST multilayers exhibited an excellent $z_{xy}T$ of 0.20 at room temperature owing to the negligible negative effects of the interfacial resistances on the thermoelectric properties. Utilizing the high-performance MCM elements, we constructed a lateral thermopile module and demonstrated power generation of 204 mW at $\Delta T$ = 152 K, which corresponds to the highest normalized power density of 0.17 mW/cm$^2$·(mm/K)$^2$ among the transverse thermoelectric modules. MCM developed in this study paves the way toward new core technologies for energy harvesting and thermal management.

## 5. Experimental section

### 5.1 Sample preparation and characterization

$SmCo_5$/BST-based ATMLs were prepared as follows. Anisotropic $SmCo_5$-type magnet disks with a diameter of 20 mm and a thickness of 0.5 mm (YX24, Magfine Corporation) and BST alloy powder with 99.9% purity and a particle size of >200 μm (Toshima Manufacturing Co., Ltd.) were employed. The demagnetized $SmCo_5$ disks and BST powder were alternately stacked and bonded using spark plasma sintering under a pressure of 30 MPa at 450 °C for 20–30 min, where the amount of BST powder per unit layer was approximately 1.05 g that was transformed to be 0.5-mm-thick after densification. Finally, the sintered stack was cut into a rectangular shape with $\theta$ of approximately 25°. To magnetize the sintered $SmCo_5$/BST multilayers, a pulse magnetic field of 8 T was applied in the direction perpendicular to the stacking plane. The plain BST slab was also prepared using the spark plasma sintering method under the same sintering condition to measure the transport properties.

The temperature dependence of $\sigma$ and $S$ of the $SmCo_5$ and BST slabs was measured using the Seebeck-coefficient/electric-resistance measurement system (ZEM-3, ADVANCE RIKO Inc.). The temperature dependence of $\kappa$ was determined through thermal diffusivity measured using the laser flush method, specific heat measured using the differential scanning calorimetry, and density measured using the Archimedes method. The magnetization $M$ curve of $SmCo_5$ was measured via the superconducting quantum interference device vibrating sample magnetometry using Magnetic Properties Measurement System (MPMS3, Quantum Design Inc.).

The elemental maps of the cross section of the $SmCo_5$/BST multilayer were observed by SEM-EDX using Cross-Beam 1540ESB (Carl Zeiss AG). To do this, the surface of the sample was mechanically polished in advance.

The interfacial electrical and thermal resistances of the $SmCo_5$/BST multilayer were characterized as described below. The $SmCo_5$/BST multilayer sample with $\theta = 0°$ and dimensions of 3.2 × 11.2 × 1.9 mm was prepared. The position dependence of the four-terminal resistance was measured using the resistance distribution measuring instrument (Mottainai Energy Co., Ltd.), where the contact probe was moved in 10-μm increment and alternating current with an amplitude of 100 mA was applied in the stacking direction. The LIT measurements were performed using Enhanced Lock-In Thermal Emission (ELITE, DCG Systems G.K.) at room temperature and atmospheric pressure. The sample was fixed on a plastic plate with low thermal conductivity to reduce heat leakage due to thermal conduction. To enhance the infrared emissivity and ensure uniform emission properties, the top surface of the sample was coated with an insulating black ink having an emissivity higher than 0.94 (JSC-3, JAPANSENSOR Corp.). The viewing areas of the thermal images in Fig. 3f and Fig. S1 (ESI†) are 1.54 × 1.74 mm and 7.68 × 3.84 mm, respectively.

The $SmCo_5$/BST multilayer sample with $\theta = 25°$ and dimensions of 12.4 × 8.3 × 1.1 mm was prepared for the direct measurements of $S_{xy}$ and $\sigma_{xx}$. The 8.3 × 1.1 mm surfaces were covered with Cerasolzer #297 (Kuroda Techno Co., Ltd.) using the ultrasonic soldering technique to form electrodes for applying a uniform current. The copper wires were connected to the electrodes by Cerasolzer #186 (Kuroda Techno Co., Ltd.) using the soldering iron. This sample was bridged between two anodized Al blocks, one of which is connected to chip heaters and the other to a heat bath to generate $\nabla T$ in the 8.3 mm direction. A central part of the 12.4 × 8.3 mm surface was covered with a black ink and the temperature distribution was measured with the infrared camera. The two Al-1%Si wires were directly connected to the 12.4 × 8.3 mm surface with a distance of 6.5 mm to measure the ac resistance and dc voltage. Whereas the side surfaces were fully covered with a solder to have an alternating current uniformly input to MCM, $V$ and the ac resistance were measured between the point contacts inside MCM to exclude the boundary effect, which might cause the decrease in $S_{xy}$ and increase in $\sigma_{xx}$ compared with the analytical values.[26,28] The ac resistance was measured by a battery internal resistance tester (BT3562A, Hioki E.E. Corp.) with applying an alternating current with an amplitude of 100 mA. The dc voltage under the application of $\nabla T$ was measured by a nanovoltmeter (2182A, Tektronix, Inc.).

### 5.2 Construction of MCM-based thermopile module

The rectangular blocks of $SmCo_5$/BST-based ATML with $t = 0.5$ and $\theta = 25°$ were sliced into a rectangular shape with dimensions of 15.4 × 7.3 × 1.5 mm. To create electrodes to electrically connect the MCM elements in series, the 7.3 × 1.5 mm surfaces were firmly covered with Cerasolzer #297 (Kuroda Techno Co., Ltd.) using the ultrasonic soldering technique. The 14 MCM elements were alternately stacked with the opposite $\theta$ and intermediated by 0.05-mm-thick insulating paper towels and heat-resistant glue (Duralco NM25, Cotronics Corp.). After curing for more than 4 h, the neighboring MCM elements were electrically connected to form a zigzag circuit by Cerasolzer #186 (Kuroda Techno Co., Ltd.) using the soldering iron. At the ends of the thermopile circuit, enameled copper wires were connected using Cerasolzer #186 for power generation measurements. Finally, the heat transfer surfaces and the base of the copper wires were covered with the same heat-resistant glue to obtain smooth surfaces and fix the wires. The module shown in Fig. 4b was magnetized by a pulse magnetic field of 8 T along the direction of the heat current.

### 5.3 Transverse thermoelectric generation measurements

A custom-made sample holder was used to measure the transverse thermoelectric power generation. The MCM-based module was sandwiched between a heater plate and aluminum heat sink. Ethylene glycol cooled at 273 K was circulated inside the heat sink during the measurements. To form uniform thermal contacts between the heater, module, and heat sink, 2-mm-thick AlN ceramic plates and 1-mm-thick insulating polymer sheets (MANION-SC, Sekisui Polymatech Co., Ltd.) with the high thermal conductivity of approximately 25 W/mK were inserted and screwed down, where the module was in direct contact with the polymer sheets (Fig. 4c). The side surface of the module was covered with insulating black ink to measure $\Delta T$ using an infrared camera. After the application and stabilization of $\nabla T$ for 10 min, the $I_{load}$ dependence of $V$ was measured three times and averaged. To characterize the internal resistance of the MCM-based module, the $I_{load}$–$V$ curve was measured from -100 to +100 mA. To characterize the power generation performance, the $I_{load}$–$V$ curve was measured from 0 to +6000 mA. The raw data of the $I_{load}$–$V$ curves for the MCM-based module included the voltage drop due to the copper wires. Thus, the $I_{load}$–$V$ curves only for two copper wires, whose ends were short-circuited using Cerasolzer #186, were subtracted from the raw data to evaluate the pure thermoelectric performance of the MCM-based module.

## Author contributions

F.A. and K.U. conceived the idea, planned and supervised the study, designed the experiments, prepared the samples, developed an explanation of the results, and prepared the manuscript. F.A. and T.H. collected the data of the thermoelectric properties and LIT images. F.A. calculated the transverse thermoelectric properties. H.S.A. observed and analyzed the microstructure. F.A. and Y.I. analyzed the interfacial electrical resistance. A.A. and H.N. analyzed the interfacial thermal resistance. All the authors discussed the results and commented on the manuscript.

## Conflicts of interest

The authors declare no conflict of interest.

## Data availability

The data supporting this article have been included as part of the ESI.†

## Acknowledgements

The authors thank K. Suzuki and M. Isomura for technical supports and Y. Oikawa for valuable discussions. This work was supported by ERATO “Magnetic Thermal Management Materials” (No. JPMJER2201) from Japan Science and Technology Agency (JST), Grants-in-Aid for Scientific Research (KAKENHI) (No. 24K17610) from Japan Society for the Promotion of Science (JSPS), and NEC Corporation.

# Supplementary information

**Multifunctional composite magnet realizing record-high transverse thermoelectric generation**

Fuyuki Ando,*[a] Takamasa Hirai,[a] Abdulkareem Alasli,[b] Hossein Sepehri-Amin,[a] Yutaka Iwasaki,[a] Hosei Nagano[b] and Ken-ichi Uchida*[ac]

[a.] National Institute for Materials Science, Tsukuba 305-0047, Japan

[b.] Department of Mechanical Systems Engineering, Nagoya University, Nagoya 464-8603, Japan

[c.] Department of Advanced Materials Science, Graduate School of Frontier Sciences,

The University of Tokyo, Kashiwa 277-8561, Japan

*E-mail: ANDO.Fuyuki@nims.go.jp; UCHIDA.Kenichi@nims.go.jp

### Note S1. Observation of nonuniform charge-to-heat current conversion in MCM

To buttress the appearance of the transverse thermoelectric conversion, we measured the charge-current-induced temperature modulation in $SmCo_5$/BST-based MCM by using the LIT technique. The application of $\mathbf{J}_c$ induces a net heat current $\mathbf{J}_q$ in the orthogonal direction by the off-diagonal Peltier effect, which is the Onsager reciprocal of ODSE.[1] We performed LIT measurements at $J_c = 1$ A and various $f$ values in the cross-sectional and top-side configurations, as shown in Fig. S1a,b, and obtained $A$ and $\varphi$ images using Enhanced Lock-In Thermal Emission (ELITE, DCG Systems G.K.) at room temperature and atmospheric pressure. The $SmCo_5$/BST-based MCM with $\theta = 25°$ and dimensions of 2.2 × 11.7 × 2.0 mm was used for the LIT measurements. The procedures of the mounting and surface coating of the sample were the same as the experiments in Fig. 3 and Experimental section in the main text.

Fig. S1c–f shows the results of the LIT measurements. In the cross-sectional configuration, the $A$ and $\varphi$ images at $f = 10.0$ Hz indicate that alternating heating and cooling signals are localized near the $SmCo_5$/BST interfaces in a similar manner to that shown in Fig. 3f in the main text. Importantly, the magnitude of the $A$ signals for $\theta = 25°$ was nonuniform along the oblique interfaces because the charge current flowed nonuniformly due to the anisotropic electrical conductivity; this is the origin of transverse thermoelectric conversion by the off-diagonal Peltier effect. These local heating and cooling signals were broadened through thermal diffusion as $f$ decreased. The $A$ image at $f = 0.1$ Hz, showing a nearly steady-state amplitude of the temperature modulation, indicates that large temperature changes occurred near the upper and lower edges of the sample. The $\varphi$ signals varied from approximately 0° (red regions) to 180° (blue regions) in the direction perpendicular to $\mathbf{J}_c$, indicating transverse thermoelectric heating and cooling, respectively.[1] This behavior is more evident in the results for the top-side configuration. Fig. S1e,f shows that almost uniform $\varphi$ signals of approximately 180° are induced by the longitudinal charge current at $f = 0.1$ Hz, confirming that $SmCo_5$/BST-based MCM operates as a transverse thermoelectric converter. Fig. S1g shows the line profiles of the average temperature modulation signals in the area defined by the white dotted rectangles in Fig. S1e,f, where the $A$ signals periodically change due to the multilayer structure. Our MCM exhibited a considerably larger temperature modulation of 1–2 K at $f = 0.1$ Hz than that in the previously reported $Nd_2Fe_{14}B/Bi_{88}Sb_{12}$-based ATMLs (approximately 0.2 K) with the same $J_c$ and $f$ values.[1] The observation of the large off-diagonal Peltier effect ensures that our MCM exhibits large transverse thermoelectric generation by ODSE.

## Note S2. Influence of magnetic state on intrinsic thermoelectric performance of MCM

The magnetic state of $SmCo_5$ does not contribute to the transverse thermoelectric generation performance in our MCM-based module for the following reasons. Based on the transport measurement results in the previous reports,[1,2] the modulation of ODSE through the magneto-Seebeck and magnetoresistance effects in $SmCo_5$ and BST was estimated to be less than 0.6%. Because the electric fields induced by the ordinary Nernst effect in BST and ANE in $SmCo_5$ can appear only in the cross-product direction of the *x*- and *y*-axes due to the strong magnetic anisotropy of $SmCo_5$ in the stacking direction (Fig. 2a in the main text), they are not superposed on the thermopower in the *x*-axis.

## Note S3. Calculation of thermoelectric generation performance of MCM-based module

We calculated the $\Delta T$ dependence of $V_{oc}$, $R_{module}$, $P_{max}$, and conversion efficiency $\eta$ of our MCM-based module using the temperature dependence of analytically calculated transverse thermoelectric properties in Fig. S3. Snyder's method[3,4] was adopted to accurately calculate $\eta$ for transverse thermoelectric conversion. By introducing the device figure of merit $Z_{xy}T$ of a transverse thermoelectric generator, $\eta$ is expressed in the traditional manner as

$$\eta = \frac{\Delta T}{T_h}\frac{\sqrt{1+Z_{xy}T}-1}{\sqrt{1+Z_{xy}T}+T_c/T_h} \quad (1)$$

where $\Delta T/T_h = (T_h - T_c)/T_h$ is the Carnot efficiency with $T_{h(c)}$ being the hot (cold) side temperature and the remaining factor is the reduced efficiency depending on $Z_{xy}T$. The essential difference between $z_{xy}T$ and $Z_{xy}T$ is that whereas $z_{xy}T$ is defined by $S_{xy}(T)$, $\sigma_{xx}(T)$, and $\kappa_{yy}(T)$ at specific $T$, $Z_{xy}T$ (and $\eta$) is calculated from the temperature dependence of $S_{xy}(T)$, $\sigma_{xx}(T)$, and $\kappa_{yy}(T)$ from $T_c$ to $T_h$ and variable with the applied load current. To exactly calculate $\eta$ using $Z_{xy}T$, the temperature-dependent thermoelectric properties with an inclement of 1 K were obtained by the linear interpolation and input to the spreadsheet given in the literature.[4] Then, by taking $T_h$ and $T_c$ measured by the infrared camera and the dimensions and number of the MCM-elements into account, the $\Delta T$ dependence of $V_{oc}$ and $R_{module}$ was analytically calculated as shown in Fig. S5a,b. The relative current density $u$, which refers to the applied load current divided by the input heat current, was optimized to get the maximum output power $P_{max}$ value at each $\Delta T$ value for the comparison with our experiment. Fig. S5c shows the result of the analytical calculation for $P_{max}$ with the experimental values. The $\Delta T$ dependence of the calculated $\eta$ under the same $u$ value is shown in Fig. S5, where the $\eta$ value at $\Delta T$ = 152 K is estimated to be 2.4% (without any losses) and 1.6% (by multiplying the loss fraction of the measured $P_{max}$ to the calculated one in Fig. S5c) in maximum. The corresponding normalized conversion efficiency, where $\eta$ is divided by the Carnot efficiency for the fair comparison of the thermoelectric generation performance, is in the range of 5.2−7.6% in our MCM-based module. The value is much higher than that of Bi/Cu- and Ni/BST-based ATML modules (3.9% and 3.6%, respectively),[5,6] but still lower than that of the longitudinal thermoelectric modules[7–13] mainly due to the high thermal conductivity of our MCM. Note that the loss fraction of $P_{max}$ can be suppressed simply by elongating the MCM element in the **E** direction because the shunting effect at the boundaries to the electrodes predominantly decreases $V_{oc}$ and hence $P_{max}$.[14,15] Thus, the developments in both the materials performance and device engineering are significant for further improvement in $\eta$ and thermoelectric applications of MCMs.

### Note S4. Efficient energy harvesting utilizing multifunctionality

To demonstrate technological advantages of MCM, we propose a new device concept where the multifunctionality of permanent magnet features and transverse thermoelectric conversion improves thermal energy harvesting performance even if the intrinsic material properties are unchanged. Fig. S6a shows the schematic of a cross-sectional view of the MCM-based module acting as a thermal energy harvester. The magnetic attractive force of the MCM-based module enables not only an easy installation onto a heat source made of a magnetic material but also an efficient heat input by eliminating the vacant space and reducing the contact thermal resistance between MCM and the heat source. In addition, the lateral thermopile geometry allows a surface area on the cool side to increase by changing the height of the neighboring MCM elements, enabling efficient heat release to air atmosphere without attaching a heat bath; such a configuration is suitable for thermal energy harvesting. Hereafter, we refer to the device structure depicted in Fig. S6a as a built-in heat sink (BHS).

For the proof-of-concept demonstration, we prepared the MCM-based modules with and without BHS and compared their thermoelectric power generation performance without attaching a heat bath at the cold side. In a similar manner to the experiments in Fig. 4 in the main text, the additional three thermopile modules were constructed for the control experiment: (i) a demagnetized module without BHS, (ii) a magnetized module without BHS, and (iii) a magnetized module with BHS. For (i) and (ii), 8 elements of $SmCo_5$/BST-based ATMLs with $t = 0.5$ and $\theta = 25°$ were prepared with a rectangular shape of 12.2 × 7.3 × 1.5 mm. Meanwhile, for (iii), 4 elements with two different heights (8 elements in total) were alternately connected. To keep the same volume as that of (i) and (ii), the size of the smaller (larger) element was 12.2 × 6.3 × 1.5 mm (12.2 × 8.3 × 1.5 mm). The module construction process was the same as mentioned in Experimental section in the main text. To magnetize (ii) and (iii) after the construction, a pulse magnetic field of 8 T was applied along the direction of the heat current. The internal resistances of these modules are 20.4, 21.8, and 21.0 mΩ, respectively, which confirms almost the same volume and electrode contact conditions. The photograph in Fig. S6b shows the constructed modules (i) and (iii). Fig. S6c shows the experimental setup for the thermoelectric generation measurement in an air-cooled condition. A ferromagnetic SUS plate with dimensions of 150 × 1 × 150 mm was coated by a black ink to observe $T_{SUS}$ using the infrared camera and heated by a ceramic hot plate. The three MCM-based modules (i)−(iii) were put on the SUS plate intermediated by 1-mm-thick insulating polymer sheets (MANION-SC, Sekisui Polymatech Co., Ltd.) with a thermal conductivity of ~25 W/mK. The copper wires connected to the modules were fixed by curing tapes (blue color parts in the photograph in Fig. S6c) so that the modules did not move during the measurement. The top surfaces were continuously cooled by an air flow using a personal fan (BH-BZ10/TP, Panasonic Corp.) to increase the heat transfer coefficient. After setting the temperature of the ceramic hot plate and waiting for 10 min to stabilize $\nabla T$, the four-terminal $I_{load}$-$V$ measurements were performed as the same manner in the main text.

Fig. S6d,e shows $V_{oc}$ and $P_{max}$ as a function of the surface temperature of the SUS plate $T_{SUS}$ for the three MCM-based modules. All the modules show the linear increase of $V_{oc}$ with the increase of $T_{SUS}$, which indicates that $\nabla T$ also linearly increases with $T_{SUS}$ because of the almost constant $S_{xy}$ with respect to the temperature (Fig. S3). Importantly, the $V_{oc}$ values obviously vary between the three MCM-based modules even though the intrinsic thermoelectric properties are almost the same. From the comparison between (i) and (ii), the magnetized module exhibits twice larger $V_{oc}$ than the demagnetized module, suggesting the enhancement of $\nabla T$ by reducing the thermal contact resistance between the MCM-based module and SUS plate. In addition, the comparison between (ii) and (iii) reveals that the introduction of BHS leads to the enhancement of $V_{oc}$ by ~10% owing to the efficient heat release to air atmosphere. Thus, the installation of magnetic functionality and BHS successfully contribute to increase $\nabla T$ by the efficient heat transfer between the SUS plate, MCM-based module, and air. As a result of the increase of $V_{oc}$, $P_{max}$ drastically increases (Fig. S6e). As demonstrated here, the multifunctionality of the magnetic attractive force and transverse thermoelectric conversion will bring about benefit for versatile thermoelectric applications through easy installation and efficient heat transfer.

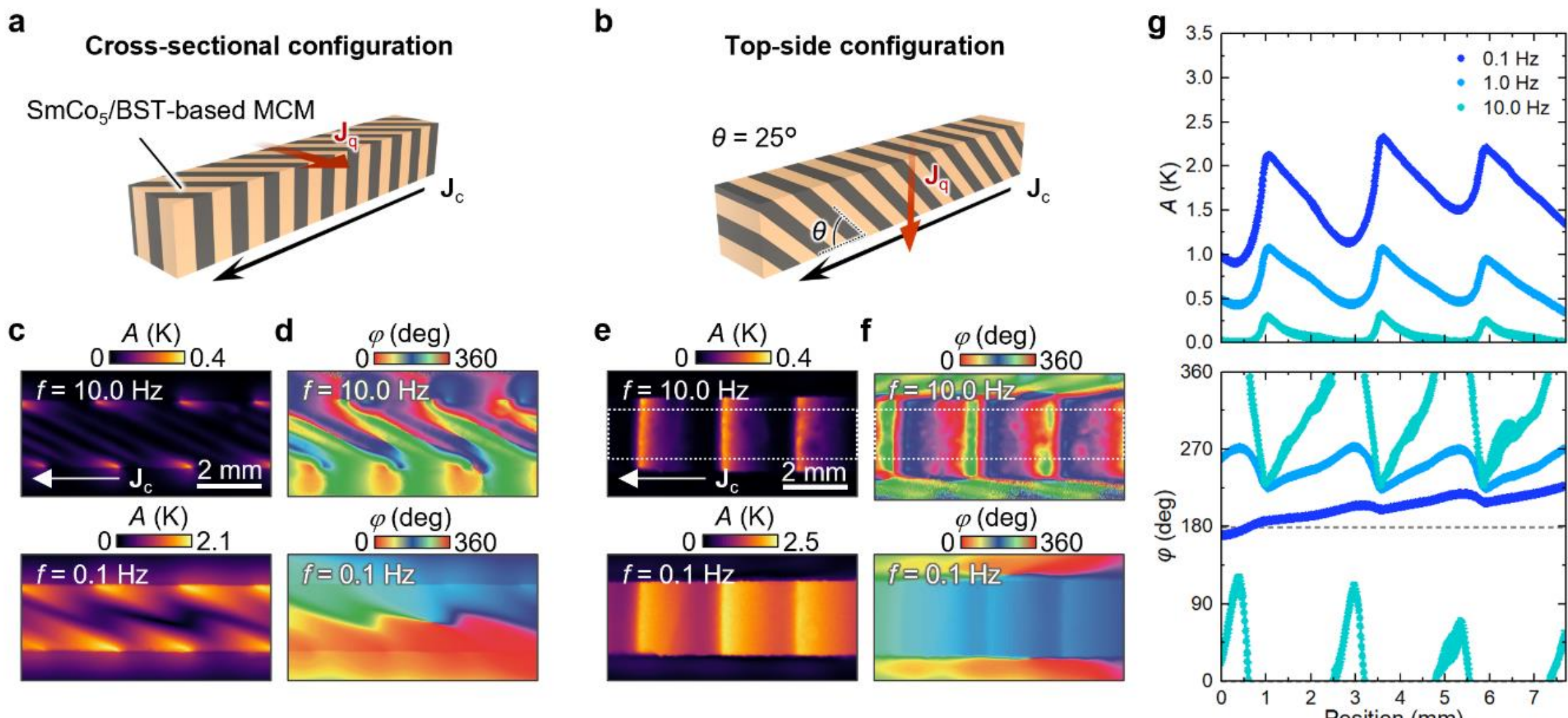

**Fig. S1** Temperature modulation due to transverse thermoelectric conversion in MCM. (a,b) Schematic of the LIT measurements for MCM with $\theta$ = 25° in the cross-sectional (a) and top-side (b) configurations. $\mathbf{J}_q$ denotes the transverse heat current generated by the off-diagonal Peltier effect in $SmCo_5$/BST-based MCM. (c,d) *A* (c) and $\varphi$ (d) images in the cross-sectional configuration at *f* = 10.0 and 0.1 Hz. *A* (e) and $\varphi$ (f) images in the top-side configuration at *f* = 10.0 and 0.1 Hz. (g) *A* and $\varphi$ profiles along the $\mathbf{J}_c$ direction at *f* = 0.1, 1.0, and 10.0 Hz for the areas defined by the white dotted rectangles with a size of 512 × 101 pixels in (e) and (f).

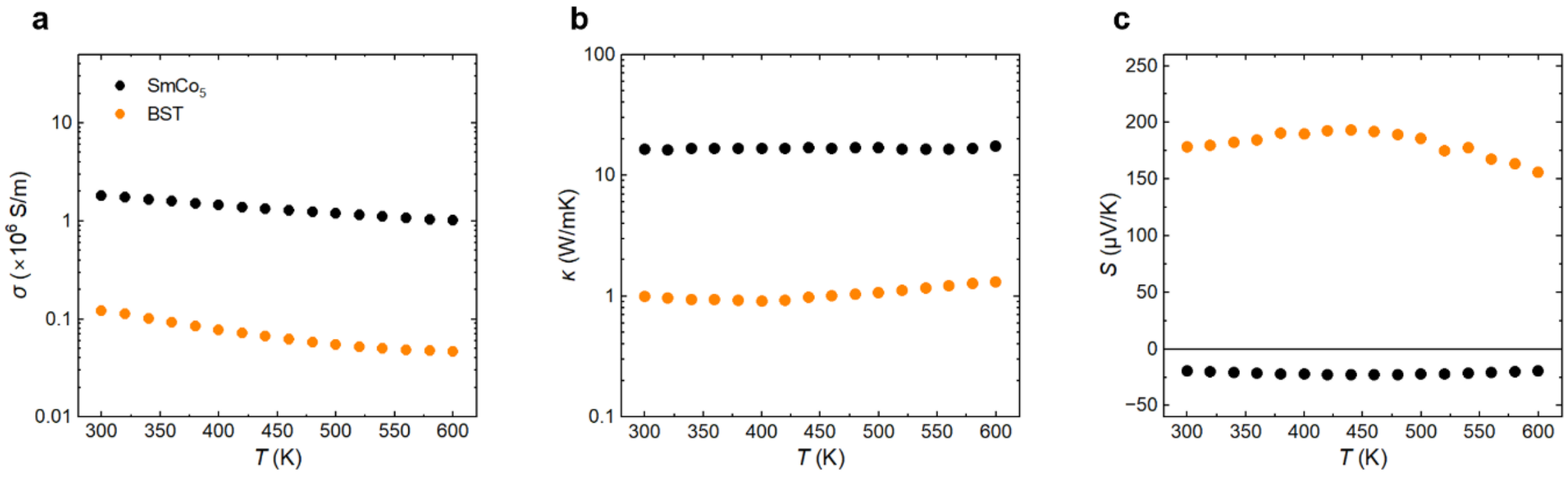

**Fig. S2** Thermoelectric properties of $SmCo_5$ and BST. (a–c) Temperature *T* dependence of the electrical conductivity $\sigma$ (a), thermal conductivity $\kappa$ (b), and Seebeck coefficient *S* (c) for the $SmCo_5$ and BST slabs.

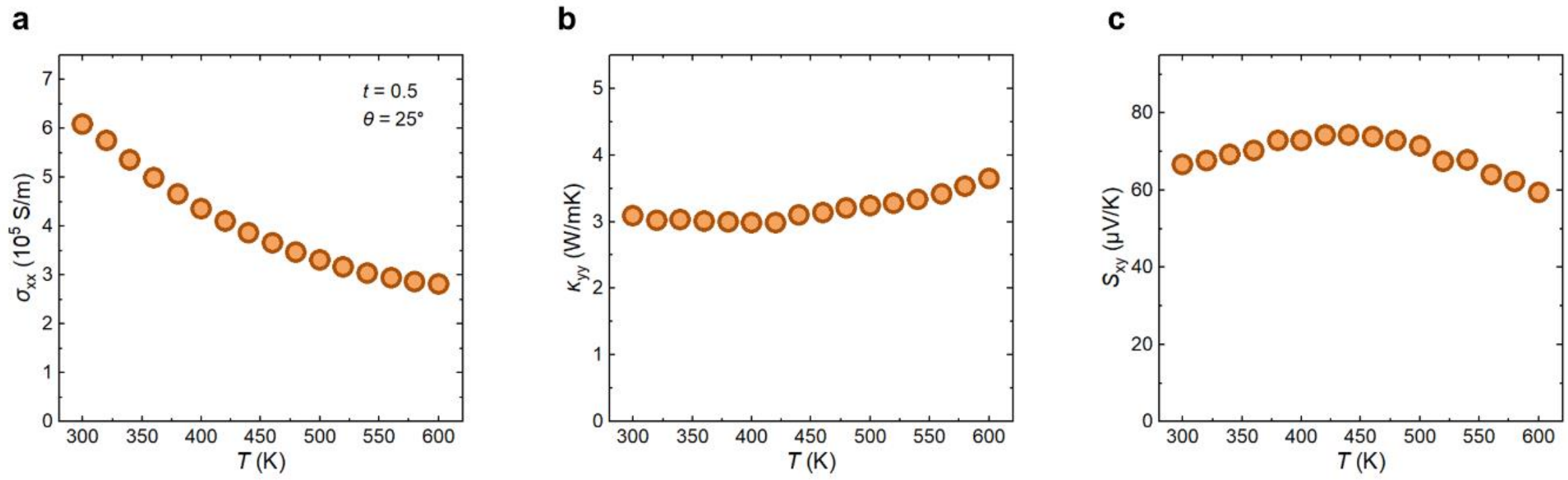

**Fig. S3** Analytical transverse thermoelectric properties of the $SmCo_5$/BST-based ATML. (a–c) $T$ dependence of the analytical transverse thermoelectric properties of the electrical conductivity in the $x$-axis $\sigma_{xx}$ (a), thermal conductivity in the $y$-axis $\kappa_{yy}$ (b), and off-diagonal Seebeck coefficient $S_{xy}$ (c) with thickness ratio $t$ = 0.5 and tilt angle $\theta$ = 25°. These parameters are calculated by substituting the experimental results in Fig. S2 into eqn (1–7) in the main text.

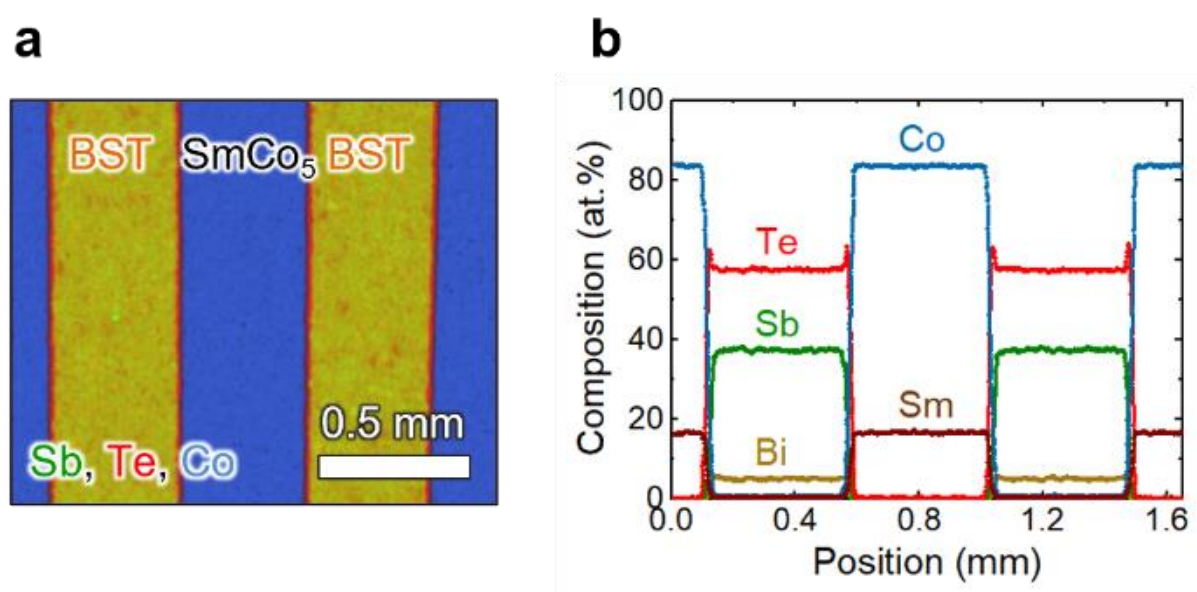

**Fig. S4** Low-magnification SEM-EDX observations. (a) SEM-EDX mapping of Sb, Te, and Co for the cross section of the $SmCo_5$/BST multilayer. (b) Line profile of the atomic ratio of Sm, Co, Bi, Sb, and Te across the stacking direction.

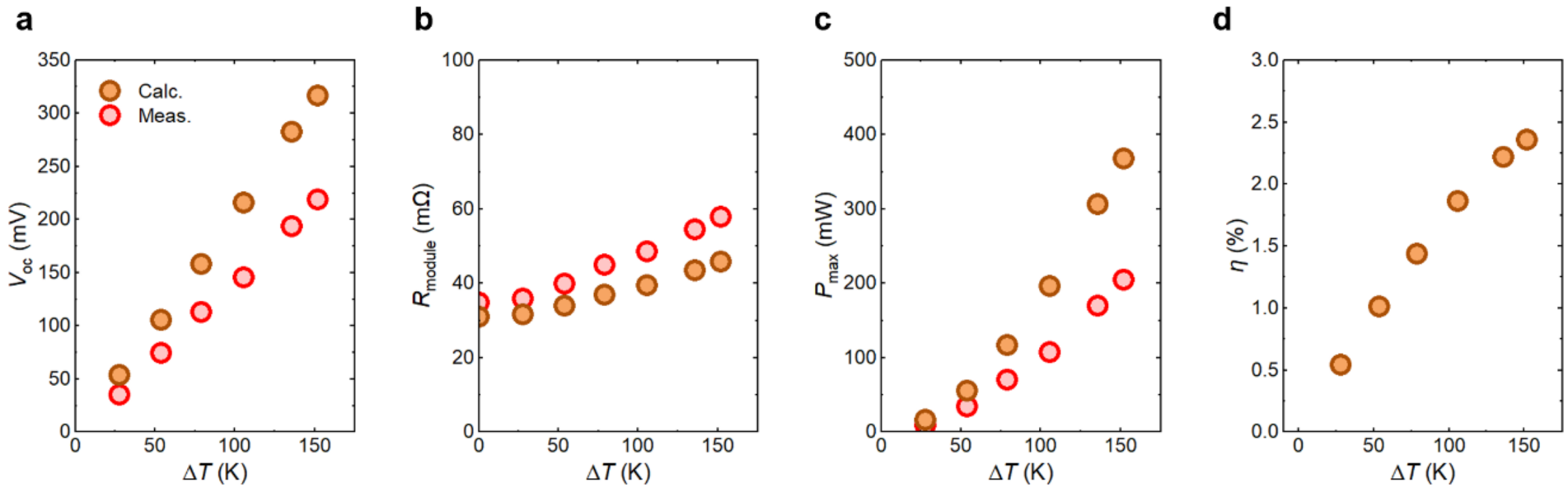

**Fig. S5** Transverse thermoelectric generation performance of MCM-based module. (a–d) $\Delta T$ dependence of $V_{oc}$ (a), internal resistance $R_{module}$ (b), $P_{max}$ (c), and conversion efficiency $\eta$ (d) for the thermopile module composed of the 14 $SmCo_5$/BST-based MCM elements with $t$ = 0.5 and $\theta$ = 25°. The orange (red) data points show the analytically calculated (experimentally measured) values. The calculated parameters are obtained by inputting the transverse thermoelectric properties based on Fig. S3 to the spreadsheet given in ref. 4.

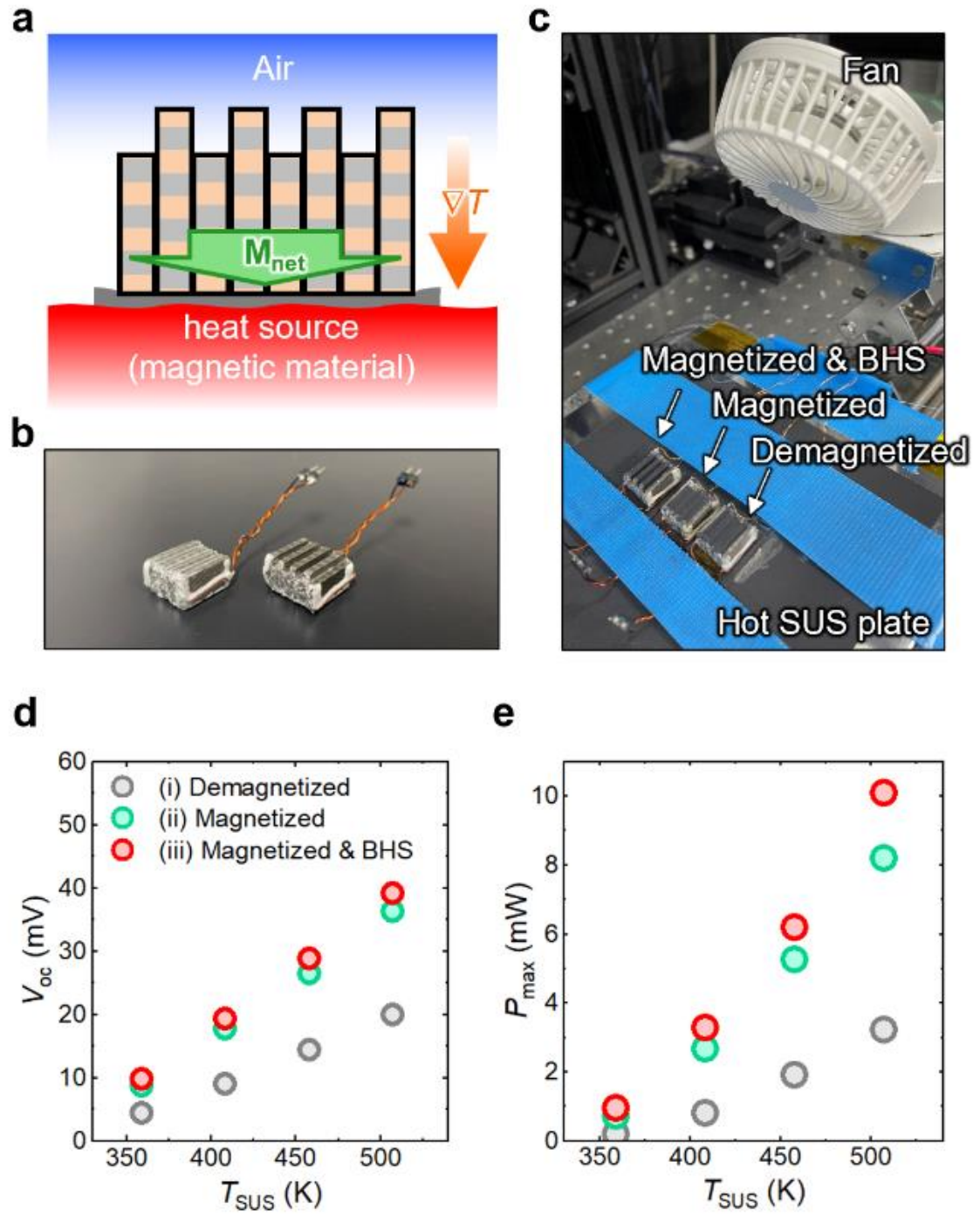

**Fig. S6** Efficient energy harvesting utilizing multifunctionality. (a) Schematic of a cross-sectional view of the MCM-based thermopile module used as a thermal energy harvester. $M_{net}$ is the net magnetization of the MCM-based thermopile module. (b) Photograph of the MCM-based modules comprising 8 elements with (right) and without (left) BHS. (c) Photograph of the measurement setup for thermoelectric power generation by (i) a demagnetized module without BHS, (ii) a magnetized module without BHS, and (iii) a magnetized module with BHS. (d,e) $T_{SUS}$ dependence of $V_{oc}$ and $P_{max}$ for (i)–(iii).